 \documentclass[final,5p,times,twocolumn]{elsarticle}

 \usepackage{graphics}

\usepackage{amssymb}

\journal{Astroparticle Physics}

\begin{document}

\begin{frontmatter}



\title{Gamma-rays from pulsar wind nebulae in starburst galaxies}


\author{Karl Mannheim, Dominik Els{\"a}sser, and Omar Tibolla}

\address{Universität W\"urzburg\\  Institut f{\"u}r Theoretische Physik und Astrophysik, 
Campus  Hubland Nord, Emil-Fischer-Str. 31, D-97084 W{\"u}rzburg, Germany\\
Corresponding author: Karl Mannheim (mannheim@astro.uni-wuerzburg.de)}

\begin{abstract}
Recently, gamma-ray emission at TeV energies has
been detected from the starburst galaxies NGC253 (Acero et al., 2009) [1] and
M82 (Acciari et al., 2009 [2]. 
It has been claimed that pion production due to
cosmic rays accelerated in supernova remnants 
interacting with the interstellar gas is responsible for
the observed gamma rays. Here, we show that
the gamma-ray  pulsar wind
nebulae left behind by the supernovae  contribute
to the TeV luminosity in a major way.
A single pulsar wind nebula produces about ten times the total luminosity of the Sun
at energies above 1 TeV during a lifetime of $10^5$~years. 
A large number of $3\times 10^4$  pulsar wind nebulae expected in a typical starburst galaxy at a distance of
4 Mpc can readily  produce the observed TeV gamma rays.

\end{abstract}


\end{frontmatter}


\section{Introduction}
\label{}

Supernova ejecta plowing through the interstellar gas form shock waves which have long been suspected as being responsible for the acceleration of cosmic rays [1-3]. The energetic cosmic ray particles traveling through the interstellar medium in a random walk can tap the shock wave energy by repeated scatterings on both sides of the shock in the diffusive-shock acceleration process. This has prompted the interpretation that the observed gamma rays are due to cosmic rays interacting with local interstellar gas and radiation [4].  However, a closer look at our  Milky Way galaxy shows the importance of sources of a different origin for the total gamma ray luminosity at very high energies, and this poses the question of their contribution to the observed gamma ray emission from starburst galaxies. \\

A scan of the inner Galaxy performed with the H.E.S.S. array at TeV energies [5] revealed the striking dominance of pulsar wind nebulae (PWNe), although some faint diffuse emission could also be detected. Studies of the total (i.e. due to sources and diffuse) gamma ray emission from the Galaxy with Fermi-LAT show a flattening above 10 GeV energies due to the increasing contribution from 
unresolved sources with rather hard spectra [6]. Late-phase PWNe show weak X-ray emission but bright TeV emission up to ages of $10^5$~years, as determined from the spin-down power of the pulsars [7].   The X-ray emitting electrons have shorter lifetimes than
the electrons producing the TeV gamma rays by  inverse-Compton scattering.  The gamma-ray lifetime is finally terminated by
adiabatic losses, breakup and diffusion into the interstellar medium [8].   
This {\em ancient PWN} paradigm provides the most elegant  solution of
the riddle of TeV sources lacking X-ray counterparts such as HESS J1507-622, HESS J1427-608 and HESS J1708-410.   A PWN
toy model has been shown to explain the salient observational features of the off-plane source HESS J1507-622 [9]. Many unidentified sources have already been identified as PWNe after their discovery (such as HESS J1857+026 or HESS J1303-631); many other unidentified sources are considered to be very likely PWNe (such as HESS J1702-420);
and also in sources that have several plausible counterparts, the PWNe contribution cannot be avoided (such as  hot spot B in HESS J1745-303 or HESS J1841-055). \\

In the following, we compare the gamma-ray  luminosities associated with PWNe and cosmic rays, respectively, in star-forming
galaxies.

\section{ PWN luminosity at TeV energies}

Prior to the Fermi era, a total of 60 galactic PWNe were known in X-rays and TeV gamma rays; according to [10] 33 of them have measured TeV fluxes. Future observations will have to confirm some of the more controversial identifications reported in [10].
Significant progress for a number of sources  has recently been achieved by the 
MILAGRO, VERITAS, MAGIC, and HESS collaborations and is ongoing.
Moreover, new TeV PWNe surrounding known pulsars have been discovered [11]. 
Fermi is now probing deeper into the population of galactic sources at GeV energies, confirming pulsars in some of the suspected PWN.  In those cases where the pulsar is not found, it could already have spun down or emit preferentially into a solid angle off the line of sight,
still in line with a PWN association. The  properties of the subsample of 28 PWNe from [10] 
which are younger than
\begin{equation} 
t_{\rm cool}({\rm 1~TeV}) = 1.3\times 10^5\  (\rm B/10\ \mu \rm G)^{-2}~{\rm years}
\end{equation}
have been considered here as representative for the putative
PWNe population in starburst galaxies.  Their average luminosity is given by $\bar L = 2.75\times 10^{34}~\rm erg~s^{-1}$
 and their average differential photon index by  $\Gamma=2.3$ ($dN/dE\propto E^{-\Gamma}$) as shown in 
Fig.~1. If the decrease of the counts at low luminosities is due to the limited flux sensitivity of the TeV observations,
$\bar L$ would be overestimated only within factors of order unity for reasonable assumptions.
Observationally, only pulsars with a spin-down
power of $\dot E_{\rm c}>4\times 10^{36}$~erg~s$^{-1}$ develop detectable PWNs [12]. 
The spin-down power {\em at birth}  is higher than $\dot E_{\rm c}$  for
canonical rotation energies of $10^{49}$~erg [13]. The high PWN luminosities are a natural consequence of this scenario.
\begin{figure*}
 \includegraphics[width=14cm]{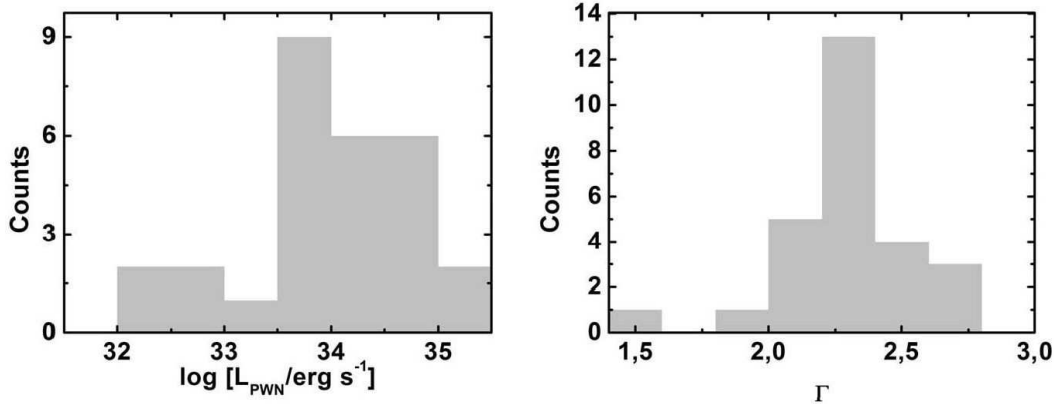}
\centering
\caption{Distribution of (1-10)~TeV luminosities and photon indices  of 28 PWN with TeV detections and
spin-down ages less than $10^5$~years from the sample compiled in ref.~[10].  Note that the
photon index does not correlate with luminosity.  }
\end{figure*}

The dominance of the PWN population on the TeV sky relates to the fact that they show hard spectra and long TeV  life-times 
compared with shell-type supernova remnants.  As long as the pulsar winds remain enclosed by the preceding supernova bubble,
their properties will be largely independent on the interstellar medium.  For high kick velocities, the PWN evolution
into the interstellar medium in starburst galaxies with their higher density and stronger magnetic field might be somewhat altered,
but we ignore this aspect here and leave the details of this problem to future work and further observational constraints.  \\

The number of PWN can be coarsely estimated from the core-collapse supernova rate.
Measurements of $^{26}$Al in the MeV range are consistent with a supernova rate of $R=0.02$ per year in the Milky Way galaxy assuming a Kroupa-Scalo initial mass function [14].   The maximum number of TeV-emitting PWN in the Milky Way galaxy is  given by
\begin{equation}
N_{\rm PWN} = 2.6\times 10^3 \left(R/0.02~\rm year^{-1}\right)
\end{equation} 
if we neglect other final states that do not develop a PWN such as black holes or low-magnetic field neutron stars.    Some neutron stars might produce the PWN with a delay given by the rise time of the magnetic field that was initially submerged under the neutron star surface.   
The corresponding number of shell-type supernova remnants, assuming a life-time of $10^4$~years for them,
is given by $N_{\rm SNR}=200$, and this compares to the number of eight SNR which have been detected at TeV energies.
If the same discovery fraction of $1:25$ is applied to the PWN,   ignoring the somewhat different observational biases for the
two types of sources for the sake of simplicity,
we expect a number of 100, which is actually  close to the number of known TeV-emitting PWN plus the unidentified TeV sources.
Although these estimates must be treated with extreme caution,  they show that the scenario is at least plausible.\\

Estimating the PWN luminosity of starburst galaxies is now straightforward. Since the starburst ages are typically of the order of $10^6 - 10^7$ years [15] and thus much longer than the cooling time of the TeV electrons, the number of PWN contributing to the TeV luminosity can be obtained from the current supernova rate (assuming steady-state activity).  Multiplying the
rate with the cooling age, the luminosity $\bar L$ per PWN,  and the
differential spectrum with $\Gamma=2.3$,
the total luminosity above 1~TeV is  given by
\begin{equation}
\parindent=0cm
L_{\rm PWN} \left(>E\right) = 7\times 10^{38}\left(R\over 0.2~{/ \rm year}\right)\left(E\over 1~\rm TeV\right)^{-0.3}\rm erg~s^{-1}
\end{equation}
As shown in Fig.~2, the TeV luminosities evaluated for the starburst galaxies with the above formula are in fair agreement with the observed values.
At lower energies  ($E<1$~TeV),    
the spectral index changes by $\Delta\Gamma= -0.5$ at the cooling break, and this effect can be seen in the synchrotron radiation component [16]. The transition from the harder to the softer spectrum will  be rather broad for a realistic distribution of magnetic
field strengths. Since the observed gamma-ray spectra of the starburst galaxies require a continuation of the
spectrum with $\Gamma\approx 2.3$ towards lower energies,
additional sources such as cosmic rays seem to be required to explain their GeV luminosities.

\section{Comparison with  cosmic-ray induced gamma ray emission}

The total cosmic ray luminosity from shell-powered supernova remnants is given by 
\begin{equation}
L_{\rm CR} = 6\times 10^{41}\left(R\over 0.2~{\rm year^{-1}}\right) \left(E_{\rm SNR}\over 10^{51}~{\rm erg} \right)
\left(\varepsilon\over 0.1\right)~{\rm erg~s}^{-1}
\end{equation}
where $E_{\rm SNR}$ denotes the kinetic energy and $\varepsilon$  the acceleration efficiency. 
If the cosmic rays are efficiently stored, they loose their energy by inelastic interactions with
the interstellar medium, and the calorimetric gamma ray (and neutrino) luminosity becomes a significant
fraction of the luminosity Eq.(4) [26,27].
 In fact, cosmic ray heating of the dense star-forming clouds in starburst galaxies has
been directly observed [28], and the observed gamma ray spectra at GeV energies are indeed quite flat.
In the absence of other loss processes,  the
fraction of the cosmic ray luminosity that ends up in GeV gamma rays energies can
been determined numerically to be $\sim0.25$ [29],
 and so we expect 
$L_{\rm CR,\gamma}({\rm GeV})\simeq 1.5\times 10^{41}(R/0.2{\rm year^{-1}})(\varepsilon/0.1)~{\rm erg~s}^{-1}$
 in the calorimetric limit. 
With the mean distances from the NED 3.8 Mpc for M82 and 3.1 Mpc for NGC253
the observed luminosities are
$L_{\gamma,\rm M82}({\rm GeV})=2.2\times 10^{40}\ {\rm erg\ s^{-1}}$ and
$L_{\gamma,\rm NGC253}({\rm GeV})=5.6\times 10^{39}\ {\rm erg\ s^{-1}}$,
implying an efficiency of $\varepsilon\simeq 0.01$ for $R_{\rm M82}=0.25$~year$^{-1}$
and $R_{\rm NGC253}=0.07$~year$^{-1}$ (see caption of Fig.2 for references).  
The values
 are  lower than the values of $\varepsilon\simeq 0.1$ in the Ginzburg-Syrovatskii scenario for the origin of cosmic rays in the Milky Way galaxy
[3], perhaps due to adiabatic losses in the overpressured starburst region
or the hotter interstellar medium compared to the Milky Way.
\begin{figure*}
 \centering
\includegraphics[width=15cm]{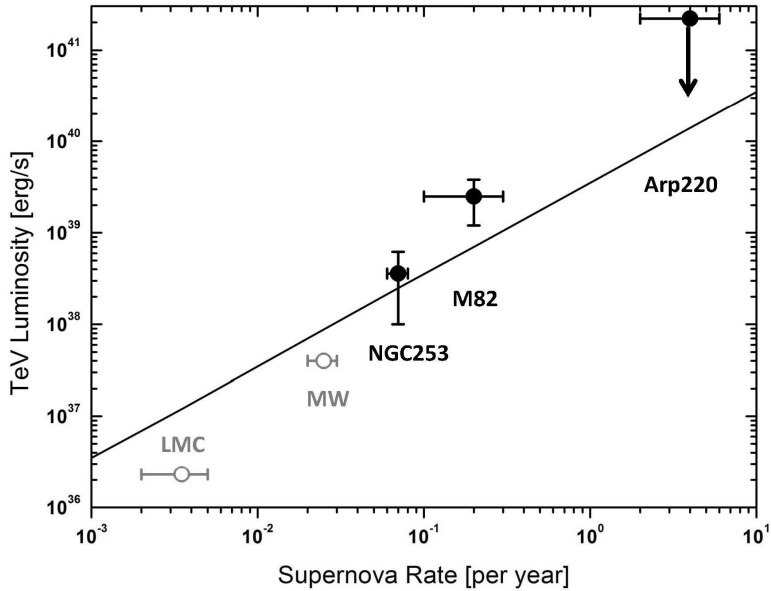}
\vskip-1cm
\caption{Comparison of the PWN luminosities at (1-10)~TeV expected from Eq.(3)  ({\em solid line}) with the observed
luminosities ({\em data points}).   The  error bars of the
{\em filled circles} reflect the uncertainties of the supernova rates [12,17-23] and of the observed TeV luminosities [1,2,24], accounting for distance and flux uncertainties.
Redshift-independent distance measurements were retrieved from the NED (nedwww.ipac.caltech.edu). 
The {\em grey open circles} denote 
the estimated TeV luminosities of the Milky Way galaxy and the Large Magellanic Cloud (dominated by 30 Doradus),
respectively by extrapolating from Fermi 
measurements with  a power law of  index -2.3 above 10~GeV [6,25].    }
\end{figure*}

Pion production energy losses further compete with advection and diffusion.  
Since the diffusive escape time $t_{\rm diff}$ decreases with increasing energy, there is
an energy $E_{\rm diff}$ at which 
the pion production time scale $t_\pi=2.5\times 10^5 (n/200~{\rm cm^{-3}})^{-1}$\ years scaled with the mean
density of the interstellar gas $n$ becomes larger than $t_{\rm diff}$.
The gamma ray luminosity in the optically thin range, i.e. at $E>E_{\rm diff}$,  is given by
\begin{equation}
L_{\rm CR,\gamma}(E)={L_{\rm CR}t_{\rm diff}(E)\over t_\pi}
\end{equation}
 In the Milky Way Galaxy, $t_{\rm diff}$ can
be determined from the $^{10}$Be/$^9$Be isotope ratio in cosmic rays, showing the energy dependence 
$t_{\rm diff}\propto E^{-0.5}$ above 1~GeV. Since the energy dependence results
from universal  properties of turbulent transport, we can assume the same energy dependence for the escape time
in starburst galaxies. 
 A similar energy dependence in starburst galaxies would mean that
the cosmic ray particles must diffuse out of the high-density star forming region and enter the low-density wind zone
at sufficiently high energies.  Here, advective transport becomes dominant which further reduces the efficiency
of the cosmic rays to produce TeV gamma rays.
The diffusion coefficient inferred from  radio measurements of starburst galaxies in the wind zone has a value of 
$D=2\times 10^{29}$~cm$^2$~s$^{-1}$ at GeV energies [30], and even larger values might be appropriate for a
 hot, turbulent, and highly magnetized star-forming region.
The transition from diffusive to advective propagation occurs at a scale height  [31] of 
\begin{equation}
z = 650~{D/2\times 10^{29}~\rm cm^2~s^{-1}\over v/1000~\rm km~s^{-1}} ~{\rm pc }
\end{equation}
considering that the wind speed reaches $\sim 1000$~km~s$^{-1}$ [32-34].   
The corresponding conservative estimate of
the diffusive escape time at GeV energies is thus given by 
\begin{equation}
t_{\rm diff} = 6\times 10^5 {( z/650~{\rm pc})^2\over D/2\times 10^{29}~\rm cm^2~s^{-1}}\left(E\over {\rm GeV}\right)^{-0.5} ~{\rm years.}
\end{equation}
The diffusion time scale becomes shorter than $t_\pi$ above $E_{\rm diff}\approx 10$~ GeV.
The steady-state spectrum due to cosmic ray interactions displays the energy dependence of the injection
spectrum $\Gamma_{\rm i}=2.2$ for  $E< 10$~GeV
\begin{eqnarray}
&L_{\rm CR,\gamma}(E)=\nonumber \\
&1.5\times 10^{41}\left(R\over 0.2~{\rm year^{-1}}\right)\left(\epsilon\over0.1\right) \left(E\over 1~{\rm GeV}\right)^{-0.2}~{\rm erg~s}^{-1}
\end{eqnarray} 
Above $E_{\rm diff} \approx 10$~GeV, the spectrum steepens  according to Eq.(5)
\begin{eqnarray}
&L_{\rm CR,\gamma}(E)=\nonumber\\
& 9\times 10^{40}\left(R\over 0.2~{\rm year^{-1}}\right)\left(\epsilon\over0.1\right) \left(E\over 10~{\rm GeV}\right)^{-2.7}~{\rm erg~s}^{-1}
\end{eqnarray} 
Due to this steepening, the PWNe
with their flatter spectrum $\bar\Gamma=2.3$ come into play towards higher energies.
Combining Eq.(3) and Eq.(8) and adopting $\varepsilon=0.01$ we obtain
\begin{eqnarray}
&\Gamma({\rm GeV-TeV})=\nonumber\\
&2+{1\over 3}\left(\log\left[L_{\rm CR,\gamma}({\rm GeV})\right]-\log\left[L_{\rm PWN}({\rm TeV})\right]\right)\simeq2.4
\end{eqnarray}
in fair agreement with the observations given the crude assumptions.
The agreement becomes somewhat better by including
the cosmic ray component from Eq.(9) at TeV energies. The result is robust, i.e. independent of the supernova rate. 
Currently, the observations are too sparse at 10 GeV to 1 TeV  to show the shallow dip that might emerge
due to the steepening of the cosmic-ray induced component and the onset of the  PWN component.

\section{Discussion and conclusions}

Starburst galaxies such as M82 and NGC253 show a much harder GeV gamma-ray spectrum than the Milky Way
galaxy, and this is in line with the high gas density in the star forming regions and a cosmic-ray origin
of the gamma rays.   However, diffusive-advective escape of the cosmic rays from the star-forming regions should
lead to a steepening of the cosmic-ray induced gamma-ray spectrum above $\approx 10$~GeV, in which case
the cosmic rays would fall short in explaining the high TeV luminosities. The effects of the escaping cosmic rays
can be seen on larger scales in the radio emission associated with the fast superwinds.
The PWNe associated with core-collapse supernovae in star-forming regions can readily explain the observed
high TeV luminosities.  
 Since the injection spectrum of cosmic rays, which determines the slope of the observed spectrum
between 100~MeV and 10~GeV, has the slope $\Gamma=2.2-2.3$ and the PWN also have a spectrum
with $\Gamma=2.3$ above $\approx 100$~GeV, the total spectrum from GeV to TeV shows little change
across a wide band. Measurements of the TeV luminosities
of galaxies such as  LMC or Arp220 will be important to better understand the relative contributions
of the gamma-ray emitting components in starburst galaxies.\\




\bibliographystyle{model1-num-names}
\bibliography{<your-bib-database>}



\section*{Acknowledgements}
We are indepted to the memory of Okkie de Jager who encouraged us to put forward this research paper. 
O.T. acknowledges support by the BMBF under contract 05A08WW1.

\end{document}